\documentclass[a4paper,english,prl,twocolumn]{revtex4}
\usepackage{CJK}
\usepackage{amsmath,amsfonts}
\usepackage{graphicx}
\usepackage{amssymb}
\usepackage{enumerate}
\usepackage{epsfig}
\usepackage{epstopdf}
\usepackage{mathrsfs}
\usepackage{dcolumn}
\usepackage{amsthm}
\usepackage{textcomp}
\usepackage[colorlinks,citecolor=blue,linkcolor=blue]{hyperref}
\usepackage{babel}
\usepackage{xcolor}
\usepackage[all]{hypcap}
\usepackage{setspace}
\newcommand{\ra}{\rangle}

\begin{document}
\title{Similar early growth of out-of-time-ordered correlators in quantum chaotic and integrable Ising chains}
\author{Hua Yan, Jiaozi Wang, and Wen-ge Wang \footnote{ Email address: wgwang@ustc.edu.cn}}
\affiliation{ Department of Modern Physics, University of Science and Technology of China,	Hefei 230026,China}
\date{\today}

\begin{abstract}
 Previous studies show that, in quantum chaotic and integrable systems,
 the so-called out-of-time-ordered correlator (OTOC) generically behaves differently at long times,
 while, it may show similar early growth in one-body systems. 
 In this paper, by means of numerical simulations, it is shown that OTOC has similar
 early growth in two quantum many-body systems, one integrable and one chaotic.
\end{abstract}
\maketitle

 \emph{Introduction}. In recent years, the so-called out-of-time-ordered correlator (OTOC) has attracted a lot of attention
 in several fields of physics, particularly high-energy physics,
 condensed matter physics, and quantum information
\cite{larkin1969quasiclassical,shenker2014black,shenker2014multiple,shenker2015stringy,
maldacena2016bound,roberts2015diagnosing,roberts2016lieb,chen2017out,fan2017out,cotler2017black,
cotler2018out,Lin2018,Kukuljan2017,scaffidi2017semiclassical,chan2019eigenstate}.
 Experimentally, it has been studied
 via nuclear magnetic resonance \cite{li2017measuring} and ion traps \cite{garttner2017measuring}.
 Study of this quantity can be traced back to an earlier work by Larkin
 and Ovchinnikov in 1969 \cite{larkin1969quasiclassical}
 in the context of superconductivity, as a measure for the instability of semi-classical
 trajectories of electrons scattered by impurities in a superconductor;
 its growth rate was found given by the classical Lyapunov exponent $\lambda_L$.
 Recently, it was proposed that OTOC may be used as a measure for quantum chaos in interacting
 quantum many-body systems \cite{kitaev2014hidden}.

 Quantitatively, denoted by $C(l,t)$, OTOC is written as
\begin{equation}
C(l,t) =  \frac{1}{2}\left\langle [W_l(t),V_0]^\dag [W_l(t),V_0] \right\rangle_\beta,
\end{equation}
 where $W_l(t)=e^{iHt}W_le^{-iHt}$ indicates the Heisenberg evolution of
 an operator $W_l$ and $\langle O\rangle_\beta \equiv \text{Tr}(e^{-\beta H}O)/\text{Tr}(e^{-\beta H})$
 indicates the initial thermal average at a temperature with $1/\beta =k_B T$.
 Here, $W_l$ represents a local operator at a site $l$ and $V_0$ is an operator
 at another site (usually fixed).

 In the specific case that $W_l$ and $V_0$  are Hermitian and unitary
 (e.g., when they are Pauli operators),
 it was found that $C(l,t) = 1-\text{Re} F(l,t)$,
 where $F(l,t) \equiv \langle W_l(t)V_0 W_l(t)V_0\rangle_\beta$, and
 grows exponentially at early times
\begin{equation}
\label{butterfly}
C(l,t) \sim  e^{\lambda_c (t-|l|/v_B)},
\end{equation}
 where $\lambda_c$ is a parameter and $v_B$ is the so-called butterfly velocity
\cite{maldacena2016bound,cotler2018out}.
 Under nature assumptions, it was found that $\lambda_c$ is bounded by  $2\pi/\beta$ in quantum systems
\cite{shenker2014black,shenker2014multiple,shenker2015stringy,roberts2015diagnosing}.
 Moreover, for local interactions in spatially extended lattice models, Eq.(\ref{butterfly})
 is not valid for long times and $C(l,t)$
 is bounded according to the Lieb-Robinson theorem \cite{lieb1972finite,roberts2016lieb},
\begin{equation}
\label{lbt}
C(l,t) \le 4||W_l||^2||V_0||^2 e^{-\mu \max\{0,|l|-v_{LR} t\}},
\end{equation}
 where $v_{LR} $ is certain parameter.

 One should note that, although $\lambda_c$ is usually called the Lyapunov exponent, it is not necessarily 
 the one that characterizes the sensitivity of chaotic motion in classical systems,
 namely, the parameter $\lambda_L$ discussed above. 
 In fact, in the kicked rotor model \cite{rozenbaum2017lyapunov,garcia2018chaos}
 and in the Dicke model \cite{chavez2019quantum},
 OTOC was found to grow as  $C(t) \sim \langle [q(t),p]^2\rangle \sim e^{2\lambda_L t}$.
 Furthermore, early growth of OTOC was found to show similar behaviors
 in some integrable and chaotic one-body systems (billiards) \cite{hashimoto2017out,rozenbaum2019quantum,jalabert2018semiclassical}.
 While, it is unclear whether OTOC may show this type of similar behavior in many-body systems.

 It is known that, for systems such as an integrable quantum Ising chain \cite{Lin2018}, 
 a Luttinger liquid \cite{dora2017out}, 
 and some models exhibiting many-body localization \cite{riddell2019out,lee2019typical}, 
 OTOC shows early power-law growths.
 But, a comparison of these early behaviors and those in quantum chaotic systems 
 has not been done. 
 In this paper, we carry out this study and compare the early growth of OTOC in two Ising chains as
 quantum many-body systems, one being integrable and the other chaotic. 
 These two chains have similar overall properties such as energy size and averaged density of states.
 This feature makes it sensible to give a quantitative comparison for the OTOC's early growths. 
 Our numerical simulations show that the two growths are very close,
 implying that the parameter $\lambda_c$ has nothing to do with whether the
 system is integrable or chaotic.

\begin{figure}
	\centering
	\includegraphics[width=1.0\linewidth]{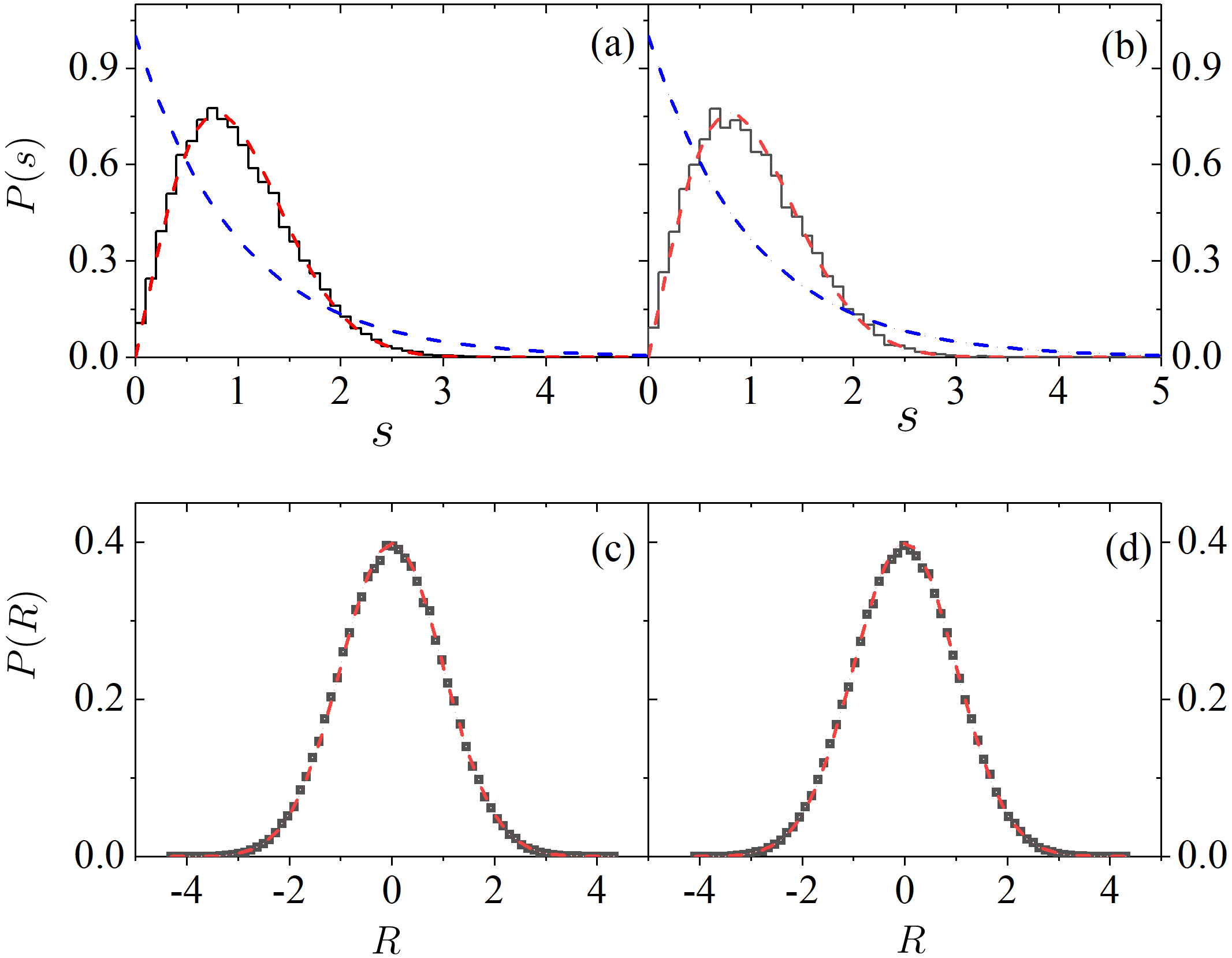}
	\caption{ (a) The nearest-level-spacing distribution $P(s)$ (histogram) of
 the defect Ising chain under the periodic boundary condition.
 The dashed line (red) indicates the Wigner-Dyson distribution
 and the dashed-dotted line (blue) represents the Poisson distribution.
 (b) Similar to (a), but under the open boundary condition.
 (c) The distribution of rescaled components $R_{\alpha i}$ of eigenfunctions in the middle energy region (squares).
 The dashed line (red) indicates the Gaussian distribution.
 Parameters: $d_1=0.5$, $d_k=1.0$, $k=6$, and $L=11$
 (the same for the following figures).}
	\label{ps_ef}
\end{figure}

 \emph{Models employed}.
 We study OTOC in two models of spin chain.
 One is the well-known quantum Ising chain in transverse field, which is integrable, with the following Hamiltonian,
\begin{equation}
H_{\rm Ising} = J_z\sum_{n=0}^{L-1}  \sigma_n^z\sigma_{n+1}^z+ h_x\sum_{i=0}^{L-1} \sigma_n^x,
\end{equation}
 and the other is a defect Ising chain,
\begin{equation}
H_{\rm defect} = d_0 \sigma_0^z + d_k \sigma^z_k + J_z\sum_{n=0}^{L-1}
\sigma_n^z\sigma_{n+1}^z+ h_x\sum_{n=0}^{L-1}  \sigma_n^x
\end{equation}
 with $k\ne 0,L$.
 We study under both the periodic boundary condition with $\sigma_{0}^z = \sigma_{L}^z$
 and the open boundary condition.
 In both Ising chains, $J_z = h_x = 1$.
 In the defect Ising chain, the parameters $d_1$ and $d_k$ are adjusted, such that the chain
is a quantum chaotic system, that is, the nearest-level-spacing distribution $P(s)$
is close to the Wigner-Dyson distribution $P_W(s) = \frac{\pi}{2}s\exp(-\frac{\pi}{4}s^2)$.
 Specifically, $d_1=0.5$, $d_k=1.0$, $k=6$, and $L=11$ [see Fig.\ref{ps_ef}].

 Recently, based on semiclassical analysis
 it was found that certain statistical property of
 eigenfunctions can also be employed as a measure for quantum chaos,
 and this is also useful in systems without any classical counterpart
 \cite{Wang_2019,wang2018characterization}.
 Let us consider eigenstates  $|\alpha\rangle$ of the defect Ising chain,
 $H_{\rm defect} |\alpha\ra = E_\alpha|\alpha\ra$,
 and their expansions in the spin configuration basis $|i\rangle$, i.e.,
 $ C_{\alpha i} = \langle i|\alpha\rangle$.
 The difference between the distribution of the following rescaled components of eigenfunctions,
\begin{equation}
\label{rescale}
R_{\alpha i} = \frac{C_{\alpha i}}{\sqrt{\langle |C_{\alpha i} |^2 \rangle}},
\end{equation}
 and the Gaussian distribution can be regarded as a measure to quantum chaos.
 Here, $\sqrt{\langle |C_{\alpha i} |^2 \rangle}$ indicates the average shape of
 the eigenfunctions.
 As seen in Fig.\ref{ps_ef}, this measure gives results in consistency with
 those given by the spectral measure discussed above.

\begin{figure}
	\includegraphics[width=1.0\linewidth]{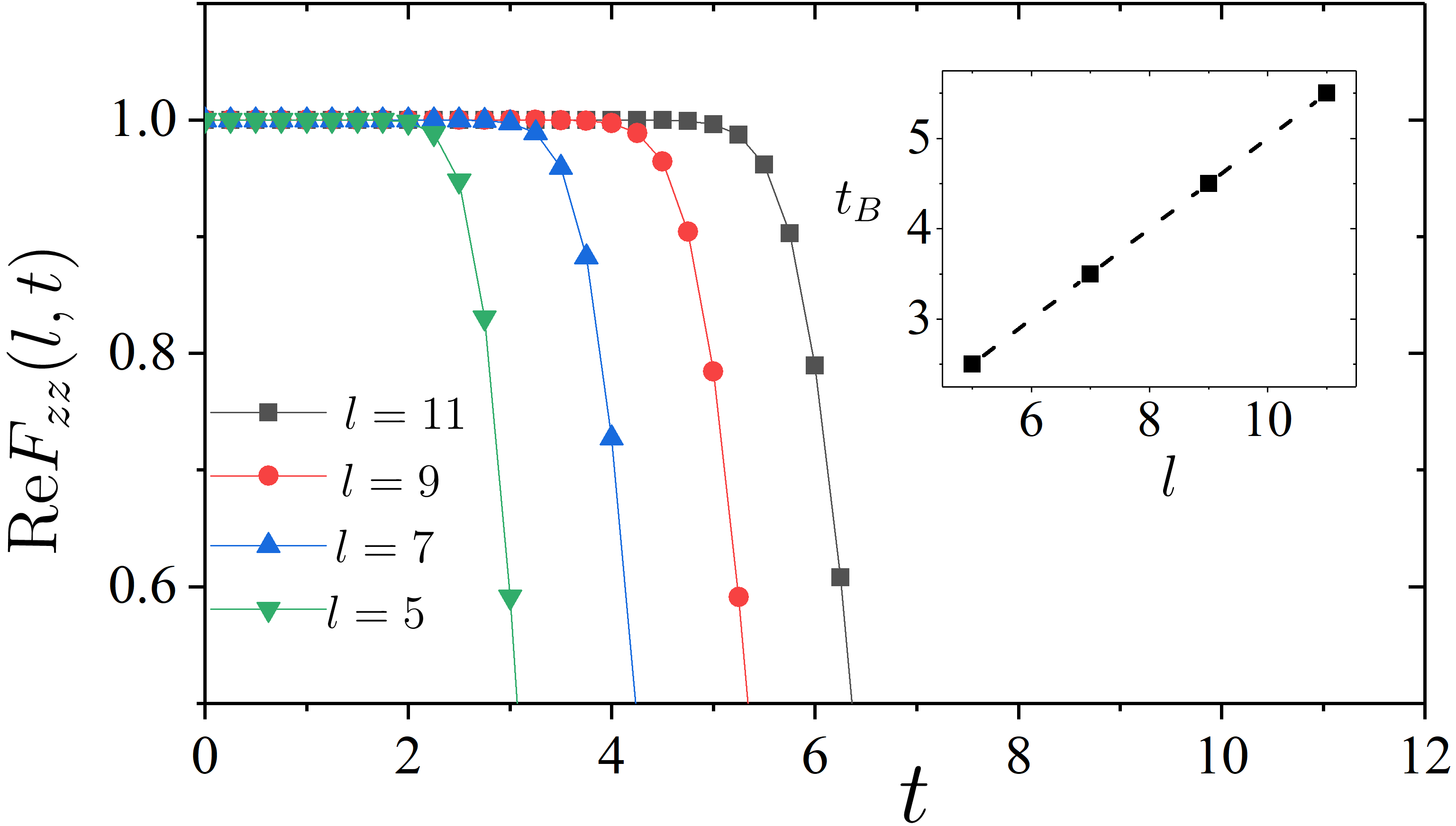}
	\caption{Variation of $\text{Re} F_{zz}(l,t) = 1-C_{zz}(l,t)$ with the time $t$
 in the defect Ising chain under the open boundary condition.
 Inset: Change of a time scale $t_B$ with the site number $l$.
 }
	\label{butterfly-vec}
\end{figure}

 \emph{Numerical simulations for OTOC.}
 Below, we first discuss early growth of OTOC in the quantum chaotic system, namely
 in the defect Ising chain with parameters given above.
 Then, we compare growths of OTOC obtained in chaotic and integrable Ising chains.

 In the computation of OTOC, the operators $W_l$ and $V_0$ are taken as
 Pauli matrices at the $l$-th site and at the first site, respectively,
\begin{gather}\label{}
 W_l = \sigma_l^\mu, \quad V_0 = \sigma_0^\nu, \quad (l \ne 0), 
\end{gather}
 where $\mu,\nu = x,y,z$ indicate directions for the Pauli operators.
 For these operators $W_l$ and $V_0$, the OTOC has a relatively simple expression, 
\begin{equation}
C_{\mu\nu}(l,t) = 1- \text{Re} F_{\mu\nu}(l,t),
\end{equation}
 where
\begin{equation}
F_{\mu\nu}(l,t) =\langle \sigma_l^\mu(t)\sigma_0^\nu \sigma_l^\mu(t)\sigma_0^\nu\rangle_\beta.
\end{equation}
 Below, we focus on the two cases of $(\mu,\nu) = (x,x)$ and $(z,z)$.
 As discussed in Ref.\cite{Lin2018}, OTOC behaves differently in these two cases in the integrable Ising chain.

 It is useful to give a brief discussion for behaviors of OTOC.
 Since $l \ne 0$, initially, $W_l(0)$ is commutable with $V_0$ and, hence, $C_{\mu\nu}(l,0) =0$.
 To get an idea about OTOC's time evolution, 
 let us consider a Baker-Campbell-Huasdorff expansion of $W_l(t)$ \cite{roberts2016lieb}, which gives
\begin{equation}\label{Wt}
W_l(t) = W_l + it[H,W_l] +\frac{(it)^2}{2!}[H,[H,W_l]]+\cdots.
\end{equation}
 For extremely short times $t$, $W_l(t) \simeq W_l$.
 With the increase of time, higher-order terms on the right-hand side (rhs) of Eq.(\ref{Wt}) should be taken into account
 in the computation of OTOC. 
 In the two spin chains discussed above, each site is coupled to its neighboring sites only. 
 Hence, for sufficiently short time $t$, $W_l(t)$ is approximately commutable with $V_0$
 and, hence, $C_{\mu\nu}(l,0) \simeq 0$.

 With further increase of the time $t$, beyond some time scale denoted by $t_B$, 
 the $l$-th term ${(it)^l}[H,\dots [H,W_l]]/ {2!}$
 on the rhs of Eq.(\ref{Wt}) becomes nonnegligible and, thus, the OTOC gets nonnegligible values. 
 Clearly, the value of $t_B$ should increase with $l$ under the open boundary condition. 
 These features can be seen in Fig.\ref{butterfly-vec}, where the early evolution of 
 $\text{Re} F_{\mu\nu}(l,t) = 1-C_{\mu\nu}(l,t)$ is plotted. 
 When Eq.\eqref{butterfly} is valid, $t_B$ may be defined by the relation $t_B = l/v_B$.
 Then, one can compute the butterfly velocity, which gives
 $v_B \simeq 2.0$ as shown in the inset of Fig.\ref{butterfly-vec},  
 in agreement with the theoretical prediction given by $v_B = 2J_z/h_x=2$ \cite{Lin2018}.

\begin{figure}
	\includegraphics[width=1.0\linewidth]{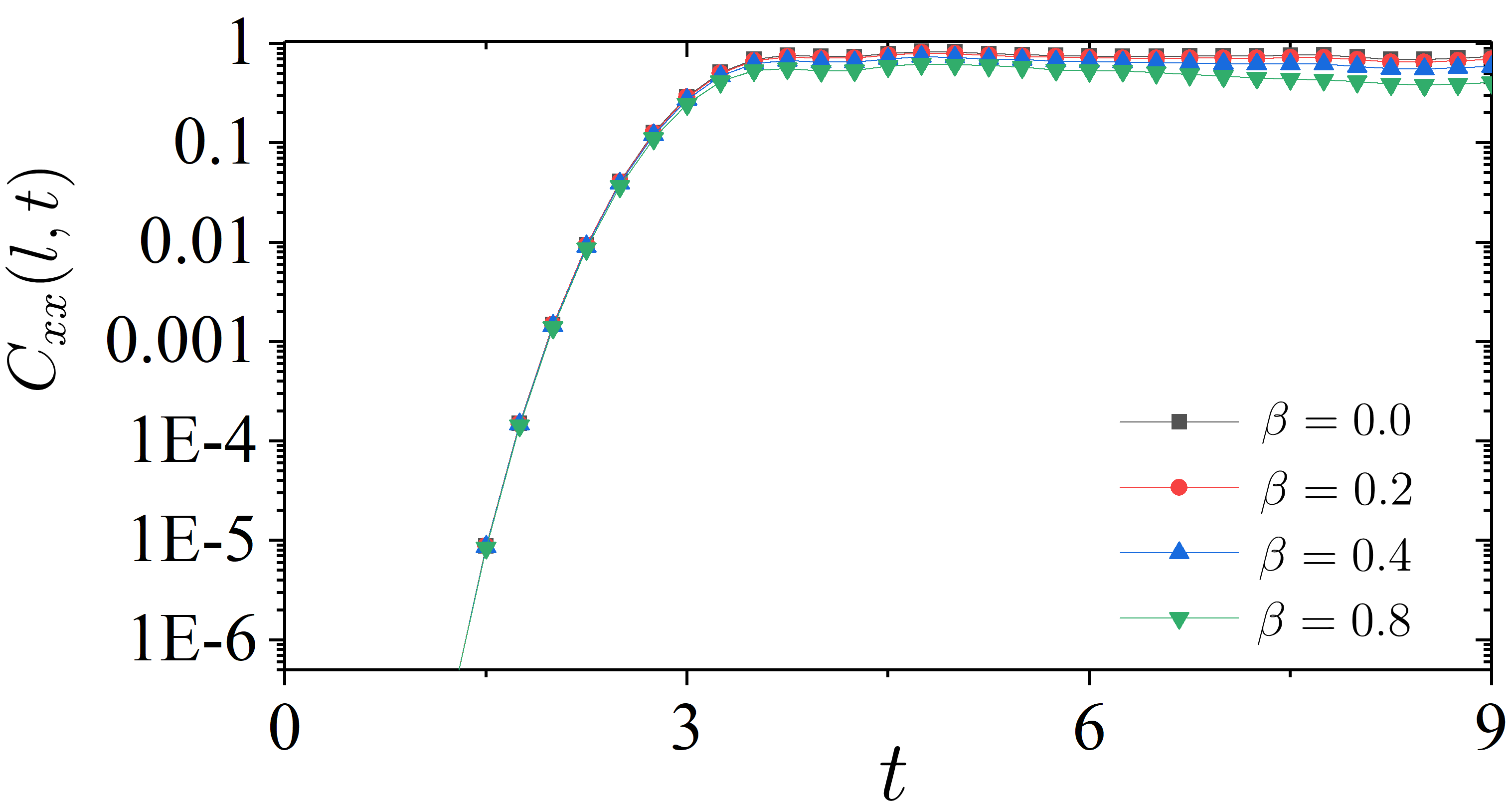}
	\caption{Variation of $C_{xx}(l,t)$ with the time $t$ at different initial temperature $\beta$,
 in the defect quantum Ising chain with $l=5$ and under the open boundary condition.
 The early increase of the OTOC is almost independent of the value of $\beta$. 
 }
	\label{otoc-te}
\end{figure}

 Numerical simulations for the dependence of the time evolution of the OTOC on the initial condition, namely,
 on the temperature $\beta$, is shown in Fig.\ref{otoc-te} for the defect Ising chain
 under the open boundary condition.
 It is seen that the early growth of OTOC is almost independent of the value of $\beta$.
 In other words, the early growth is insensitive to the initial energy
 for initial states lying in the middle energy region of the model.
 We found that a common feature of the eigenfunctions in this region of the model 
 is that they spread over almost all
 the spin configuration basis states $|i\rangle$ \cite{wang2016statistical,wang2018characterization,Wang_2019}.
 Perhaps, this wide-spreading feature plays an important role  in the early growth of OTOC.
 For long times, OTOC behaves differently depending on the value of $\beta$;
 it drops faster for larger value of $\beta$.

 Under the periodic boundary condition, the early growth of OTOC was also studied.
 The obtained results are basically similar to those
 given in the above two figures, particularly, with similar Butterfly velocity $v_B$
 but with different values of the growth-starting time $t_B$.

\begin{figure}
	\centering
	\includegraphics[width=1.0\linewidth]{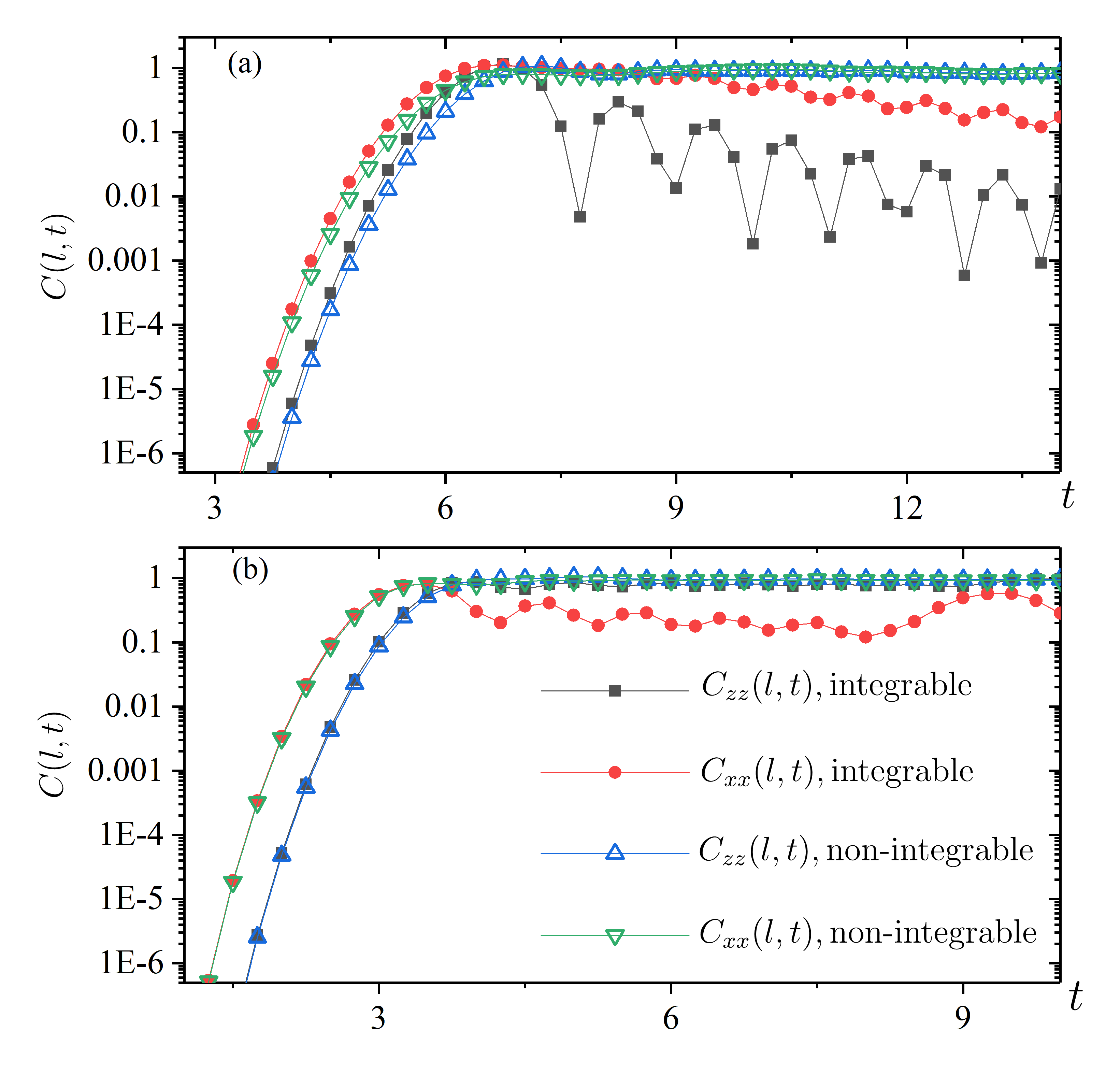}
	\caption{Comparison of early growths of OTOC in integrable and nonintegrable (chaotic) systems,
  for $(W_l,V_0) =(\sigma_l^z,\sigma_0^z)$  and  $(\sigma_l^x,\sigma_0^x)$ with $l=5$.
 (a): open boundary condition, and (b): periodic boundary condition.
}
	\label{open-closed}
\end{figure}

 Finally, we discuss a main observation of this paper.
 That is, in the two spin chains
 as quantum many-body systems, our numerical simulations show that OTOC shows similar 
 early growth in integrable and chaotic systems. 
 We have studied OTOC for two pairs of local operators, i.e., 
 $(W_l,V_0) =(\sigma_l^z,\sigma_0^z)$  and  $(\sigma_l^x,\sigma_0^x)$,
 under both the periodic and open boundary conditions (Fig.\ref{open-closed}).
 Under the periodic boundary condition, in the early-growth region,
 the values of $C_{\mu\mu}(l,t)$ ($\mu=x,z$) in the integrable Ising model
 are very close to the corresponding values in the nonintegrable (chaotic) defect Ising model. 
 For the open boundary condition, the values are also close, though not as close as in the above case.

 The above-discussed closeness of early growth of OTOC in integrable and chaotic systems
 shows that early growth of OTOC can not be employed as a measure for quantum chaos. 
 For relatively long times, the numerical results in Fig.\ref{open-closed} 
 show certain difference between integrable and chaotic systems, except for the
 squares in the lower panel which are close to the values of chaotic systems.
 To show the difference more clearly, one needs to compute for even longer times,
 at which OTOC has power-law decay in the integrable Ising chain \cite{Lin2018}.
 (See Ref.\cite{fortes2019gauging} for a recent study, which shows qualitative 
 difference between OTOC's long-time behaviors in integrable and chaotic systems.)

\emph{Concluding remarks}.
 It is found by numerical simulations that 
 the OTOC in two quantum many-body systems, one integrable and the other chaotic,
 show very close early growth under the periodic boundary condition,
 and close early growth under the open boundary condition.
 The early growth is important in quantum chaotic systems, because the OTOC approaches
 its saturation value after this early growth.

 Theoretical understanding of the above observation is still lacking. 
 One clue for future investigation is given by the following property of classical systems, that is, 
 the early motion of an integrable system, which has many degrees of freedom with incommensurable 
 frequencies, may exhibit quite irregular features. 
 In fact, it is this behavior of integrable systems that leads to the so-called Fermi-golden-rule decay of 
 quantum Loschmidt echo in integrable systems, 
 which  was first found in  quantum chaotic systems \cite{wang2010semiclassical}.

 A note. After this work was finished, 
 the authors got to know that a similar behavior of the early growth of OTOC was reported in Ref.\cite{fortes2019gauging},
 where a different chaotic Ising chain had been studied. 

 This work was partially supported by the National Natural Science Foundation of China under Grant
 Nos.~11535011 and 11775210.

\setlength{\baselineskip}{20pt}

\end{document}